\documentclass[12pt,preprint]{aastex}

\begin{document}

\title{An Extra Long X-Ray Plateau in a Gamma-Ray Burst and the Spinar Paradigm}

\author {V.Lipunov\altaffilmark{1,2,3} \& E.Gorbovskoy\altaffilmark{1,2,3}}

\altaffiltext{1}{Moscow State University}

\altaffiltext{2}{ Sternberg Astronomical Institute}

\altaffiltext{3}{Moscow Union ``Optic''}

\begin{abstract}
The recently discovered gamma-ray burst GRB 070110 displayed an extraordinary X-ray afterglow with Xray
radiation---i.e., an X-ray plateau---observed for 20,000 s. We show that the observed properties of the plateau
can be naturally interpreted in terms of the model with a spinar---a quasi---equilibrium collapsing object whose
equilibrium is maintained by the balance of centrifugal and gravitational forces and whose evolution is determined
by its magnetic field. If this model is true, then for 1 hr, the Swift X-ray telescopes recorded radiation from an
object with a size smaller than the Schwarzschild radius!                                           
\end{abstract}


\section{Introduction}


After three years of the operation of Swift space observatory (Gehrels, N., et al, 2004) it becomes evident that the
temporal behavior of many gamma-ray bursts exhibits such features as delayed flares (Chincarini, G., et al., 2007) and
early precursors (Lazzati, 2005), which can in no way be reconciled with the instantaneous point explosion model and
which are indicative of a long (compared to the duration of the gamma-ray burst) time of operation of the central
engine (Gehrels, N., et al, 2006; Wang {\&} Meszaros, 2007).

The recently discovered gamma-ray burst GRB 070110 displayed an extraordinary x-ray afterglow with x-ray radiation ---
x-ray plateau --- observed for 20~000 seconds (Troja, E., et al. 2007).The gamma-ray burst GRB 050904 exhibits a
similar behavior (Cusumano et al., 2006). Such a behavior demonstrates the central engine long-activity and gives an
insight into the mechanisms of its operation. We show that the observed properties of the plateau can be naturally
interpreted in terms of a spinar paradigm The spinar paradigm has a transparent physical meaning, which opens up a way
toward successful understanding of GRBs and the accompanying events, and allows their variety to be reduced to two
physical parameters -- initial angular momentum and initial magnetic field.

\section{Spinar Paradigm}

The prolonged activity of the central engine of gamma-ray bursts was earlier
predicted (Lipunova, (1997); Lipunova {\&} Lipunov, (1998);Vietri {\&} Stella, (1998)) to be a
result of the spinar formation. The lifetime of the spinar is determined by
the rate of dissipation of its angular momentum as a result of the
interaction of the magnetic field of the spinar with the ambient plasma.
This is the essence of the spinar paradigm. The origin of the spinar
paradigm dates back to 1960-ies when the importance of allowing for
magnetorotational effects in the process of collapse has been clearly
understood. Spinars were first invoked when analyzing the energy release by
and evolution of quasars Hoyle {\&} Fowler (1963); Ozernoy  (1966);
Ozernoy {\&} Usov, (1973)  and ejection of supernovae shells (LeBlanc
{\&} Wilson, 1970; Bisnovatyi-Kogan, 1971) .

A spinar may form in two ways: via a collision of two neutron stars
(Lipunova {\&} Lipunov, 1998) or via the core collapse of a massive star.
 Last a
situation may arise during late stages of evolution of binary systems where a binary helium star forms with an orbital
period of less than one day (Tutukov {\&} Cherepashchuk, 2003). The rate of such events is about $10^{-4} year^{-1} per
10^{11}$ $M_{\odot}$ (Bogomazov et al., 2007). This rate agrees well with the observed gamma-ray burst rate if we take
into account the narrow-beamed nature of the radiation of a gamma-ray burst. Population synthesis of a merging of
neutron stars yield a similar rate of $\sim 10^{-4}$ per year per $10^{11}$ $M_{\odot}$ for these events (Lipunov et
al., 1987).

Consider now magnetorotational collapse of a stellar core of rest mass
$M_{core}$, radius $R_{A}$ , and effective Kerr parameter (Thorne et al.,
1986) :
\begin{equation}
\label{eq1}
a_0 \equiv \frac{I\omega _0 c}{GM_{core}^2 }>1,
\end{equation}
where $I = k M_{core} R_{A}^{2}$ --- is the moment of inertia of the core; $c$ and $G$ are the speed of light and
gravitational constant, respectively, and $k$ is a dimensionless constant, which we set equal to unity for the sake of
simplicity. Parameter $a$ remains constant if the angular momentum of the core is conserved (the condition that is
evidently violated in our scenario). However, in any case, direct collapse is impossible in such a situation, because
the Kerr parameter of a black hole cannot exceed unity. Let $\alpha _{m}$ be the ratio of the magnetic energy $U_{m}$
of the core to its gravitational energy:$\alpha _m \equiv U_m /(GM_{core}^2/R)$. The total magnetic energy can be
written in terms of the average magnetic field $B$ that penetrates the spinar, $U_{m}=(1/6)B^{2}R^{3}$. In the
approximation of the conservation of magnetic flux ($BR^{2}=const$) the ratio of the magnetic and gravitational
energies remains constant throughout the collapse: $\alpha_m=const$, $U_m \propto R^{-1}$. The collapse process breaks
into several important stages (see Fig.1.). After the loss of stability virtually free-fall contraction begins with a
time scale of $t_{A }= (R_{A}^{3 }/ GM_{core})^{1/2}\sim 100 s$,
 where $R_{A} \sim 10^{11} cm$ is the initial radius of
the stellar core (Wang {\&} Meszaros, 2007 eq. 15). During the collapse the gravitational energy is hardly
radiated, but is transformed into kinetic, rotational, and magnetic energy
of the core. The rotational energy can be easily seen to grow faster than
the gravitational energy $U_{spin} \approx I\omega^2/2\propto R^{-2}$ , and
the collapse stops (Figñ. 1B) near the radius determined by the balance of
the centrifugal and gravitational forces: $\omega^2 R_B = GM_{core} /{R_B}^{2}$.
It follows from this that the initial radius of the spinar is approximately equal to
$R_B =a_0^2 R_g/2$ (here $R_g =2GM_{core}/c^{2}$ is the
gravitational radius of the core). In this case, half of the gravitational
energy is released: $E_B \approx GM^2/2{R_B}=(1/2a_0^2 )M_{core} c^2$

Because of the axial symmetry the burst must be collimated along the axis of rotation and have an opening angle of
$\Omega _B$. If $a_0^2 \le 100$ the energy of the first explosion exceeds significantly the bounding energy of the
stellar shell, and relativistic jet easily comes out. The duration of this stage is determined by the time it takes
the jet to reach the surface and by the nature of cooling, which in turn is determined by the structure of the initial
jet and the shell, and ranges from several seconds to several hundred seconds. The nature of the spectrum is determined
by the Lorentz factor of the jet (Wang {\&} Meszaros, 2007). The spinar that forms it the core interiors begins to
lose its angular momentum due to magnetic viscosity and starts to radiate its rotational energy. The angular velocity
of the spinar increases like that of a satellite whose velocity increases as it decelerates in the upper layers of the
atmosphere. The spinar contracts as its angular momentum is carried away under the influence of the maximum possible magnetic torque (Lipunov, 1992):
\begin{equation}
\label{eq2}
dI\omega /dt \approx -\int\limits_R^\infty \frac{B^2}{4\pi} r 2\pi r\,dr \approx -U_m.
\end{equation}
The time scale of the dissipation of angular momentum is $t_C \approx
I\omega /U_m =GM_{core} /c^3\alpha _m$. During this process, the
velocity of rotation of the spinar increases and the spinar luminosity not
only does not decrease, but even increases $L=-\omega  dI\omega /dt=U_m
\omega \propto R^{-5/2}$ . If computed without the allowance for
relativistic effects, the spinar light curve has the form:
\begin{equation}
\label{eq3}
L=\frac{\alpha _m }{a_0^5 }\frac{c^5}{G}(1-t \mathord{\left/ {\vphantom {t
{t_C }}} \right. \kern-\nulldelimiterspace} {t_C })^{-3/5}
\end{equation}
The luminosity remains virtually constant and equal to $L_{plato}
=\frac{\alpha _m }{a_0^5 }\frac{c^5}{G}$ while $t << tc$! Thus even in the
Newtonian approximation the spinar model predicts a plateau whose parameters
can be estimated from the latter two formulas. If the magnetic field is
sufficiently strong and $t_{C}$ is small, the spinar produces an X-ray flare.

General relativity effects begin to show up as the spinar radius approaches
the gravitational radius. In particular, the magnetic field of the collapsar
begins to vanish in full agreement with the black-hole-no-hair theorem
(Thorne et al., 1986). The general-relativity evolution for the magnetic
field of the collapsar has been computed repeatedly by several researchers
(Ginzburg {\&} Ozernoy, 1964; Kramer, 1984; Manko {\&} Sibgatullin, 1992).
The results of Ginzburg {\&} Ozernoy (1964) computations can be
approximately modified. As next formula correctly describes the behavior of
magnetic energy at large distances from the gravitational radius and yields
zero magnetic field at the event horizon:
\begin{equation}
\label{eq4}
U_m=U_0\frac{x_0 }{x}\frac{\xi (x_0)}{\xi (x)}
\end{equation}
where $\xi (x)=\frac{1}{x}+\frac{1}{2x^2}+\ln(1-1/x)$, $x=R/2R_{g}$  .

The second important group of effects consists in the reference-frame drag
in the metric of the rotating body and in relativistic effects due to the
close location of the event horizon. We use the post-Newtonian approximation
for the centrifugal force in the Kerr metric to allow for the latter two
effects (Mukhopadhyay, 2002):
\begin{equation}
\label{eq5}
g=\frac{GM(x^2-2ax^{1/2}+a^2)^2}{x^3(x^{1/2}(x-2)+a^2)^2}, \quad x=R/2R_{g}.
\end{equation}
The curve of energy release acquires the features of a burst, which can be
approximately described by the following set of elementary equations:

\begin{equation}
\label{eq6}
\begin{array}{l}
\omega^2R=g, dI\omega /dt=-U_m\\
L_\infty =\alpha ^2L_C =\alpha ^2U_m \omega, dt_\infty =dt/\alpha\\
\end{array}
\end{equation}

Here $L_\infty $ and $t_\infty $ are the luminosity and time in the
reference frame of an infinitely distant observer and $\alpha $ is the
duration function -- the ratio of the rate of the clock of reference
observers to the universal time rate at the equator of the Kerr
metric(Thorne et al., 1986):
\begin{equation}
\label{eq7}
\alpha =\sqrt {\frac{x^2+a^2-2x}{x^2+a^2}},
\end{equation}
which vanishes ( $\alpha \to 0$ ) at the horizon of the extremely rotating
black hole $R\to R_g/2$.

As the luminosity increases, the condition of shell penetration becomes satisfied at a certain time instant. A second
jet appears whose intensity reaches its maximum near the gravitational radius. In this case, the effective Kerr
parameter tends to its limiting value for a Kerr black hole $a\to 1$. The subsequent fate of the star depends on its
mass. If the mass exceeds the Oppenheimer---Volkoff limit the star collapses into a black hole. Otherwise a neutron
star forms, which continues to radiate in accordance with the magnetodipole formula $L\propto t^{-2}$ (see for example
Lipunov, 1992).

The spinar paradigm allows the observed variety of gamma-ray bursts,
precursors, and flares to be reduced to just two parameters: magnetic field
and initial angular momentum (Fig.2).

\section {X-ray plateau explanation}

What is a plateau in the spinar paradigm? The simple answer is: an extralong
X-ray plateau is an x-ray flare protracted for several thousand seconds
because of the weak magnetic field of the spinar. Figure 3à shows the
rest-frame light curve of the gamma-ray bursts GRB 070110 and GRB 050904
adopted from (Troja, E., et al., 2007). Both bursts had large redshifts
($z=2.5$ and $z=6.6$) and therefore the observer sees the duration of the
corresponding plateaux to be three-seven times longer than their rest-frame
durations. Both bursts exhibited extralong plateaux which ended abruptly at
the 8000th second. Note that the energy was computed in isotropic
approximation. We therefore did not strive to achieve the exact coincidence
of luminosities, especially because we do not know the factor of conversion
of the released energy into the x-ray flux. Moreover, we have no information
about the detailed structure of the beam pattern and therefore both bursts
can be explained in terms of the same model provided that we see them at
different angles. However, the most important factor is the duration of the
plateau.

Figures 3b-e show the exact solution of the set of equations (\ref{eq6}), which agrees excellently with the observed
plateau events for GRB 070110 and GRB 050904. Both plateaux can be best described in terms of the model of the
collapse of a $7 M_{\odot}$ star with the initial rotation energy fraction of $\alpha _m =1.0\times 10^{-7}$ and
initial Kerr parameter $a= 2.0$. With these parameter values the above scenario should appears as follows. The loss of
stability by the rapidly rotating star results in the formation of a spinar and release of energy $E_B \approx
(1/2a^2)M_{core} c^2\approx 4.5\times 10^{53}erg$. The complex process of the emergence of a high-Lorentz-factor
relativistic jet onto the surface produces an about 100-s long gamma-ray burst. After the jet comes out to the surface
an afterglow ($\sim t^{-2}$) appeares due to the curvature effect (Kumar {\&} Panaitescu, 2000). Then, after the
$300^{th}$ second, most of the energy is radiated by the bow shock, which is decelerated in the stellar wind of the
progenitor star ($t^{-1}$) (Troja, E., et al., 2007) . During this process the spinar continues to radiate at a
virtually constant luminosity, which shows up at the $1000^{th}$ second, when the afterglow faded significantly. The
luminosity remains virtually constant afterward. After a small increase of the luminosity due to the compression of
the spinar the plateau terminates with the luminosity dropping abruptly, the spinar radius becomes smaller than the
gravitational radius, and the spinar is now located inside the ergosphere of the future black hole! The abrupt
increase of the gravitational redshift and effects of the disappearance of magnetic field result in the abrupt
decrease of the spinar lumosity, which continues to fall for about 900 seconds until the intensity becomes lower than
the luminosity of the bow shock. During all this time the spinar is inside the ergosphere, $R_{g}/2 < R < R_{g}$. It
seems that mankind has never before come so close to the event horizon! Since real gamma-ray bursts are located at
large redshifts, the time intervals measured with our Earth clocks are a factor of $(1+z)$ longer than the
corresponding rest-frame time intervals, providing us with an opportunity to study the collapse inside the ergosphere
for up to $2500-9000s$, i.e., virtually during an entire hour! Figure 3b shows the synthetic light curve computed with
the allowance for the afterglow, $L=C_1 t^{-2}+C_2 t^{-1}+L_\infty $, which agrees well with the observed light curve.
The small flare at $5000-10 000s$ after the period of steep decay in GRB070110 is the result of energy input from spinar to bow shock.
After end of the plateau the level of the afterglow emission must change, but with some delay.


The model of a magnetized spinar demonstrates how the parameters of the
observed plateau depend on the physical parameters of the progenitor. Our
model naturally explains other, fainter events in all types of gamma-ray
bursts, which we discuss in a separate paper.

We are grateful to the MASTER (Lipunov et al., 2006) team
for discussions and assistance in computations. We would like to thank Dr. B.Zhang and anonymous referee for useful comments.

\clearpage

\begin{figure}
\epsscale{.45}
\plotone{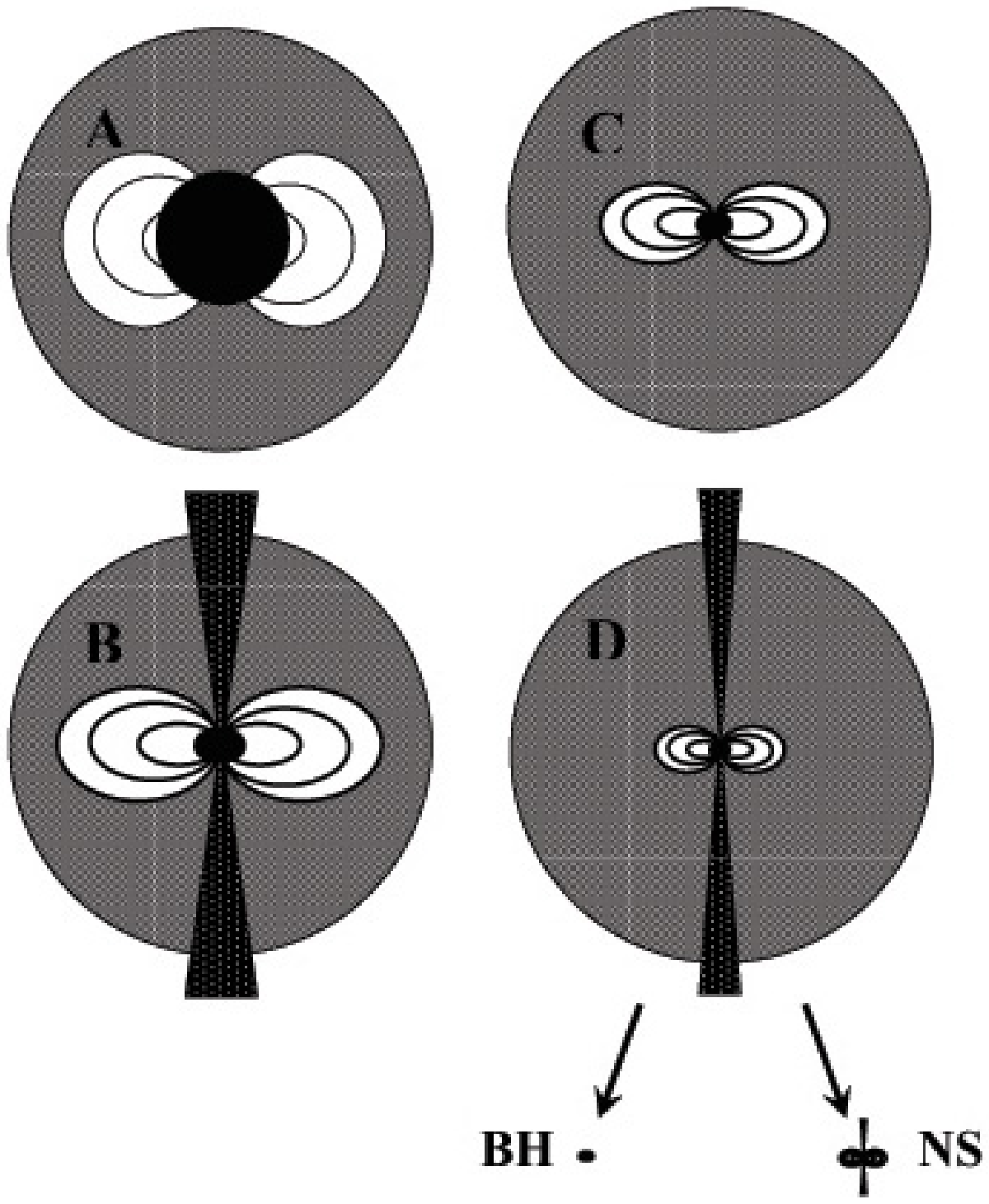}
\label{fig1} 
\caption{Schematic view of the collapse of the rapidly rotating magnetized core of a massive star. Gray
and black shaded areas show the envelope and core of the star, respectively.}
\end{figure}

\clearpage

\begin{figure}
\epsscale{0.6}
\plotone{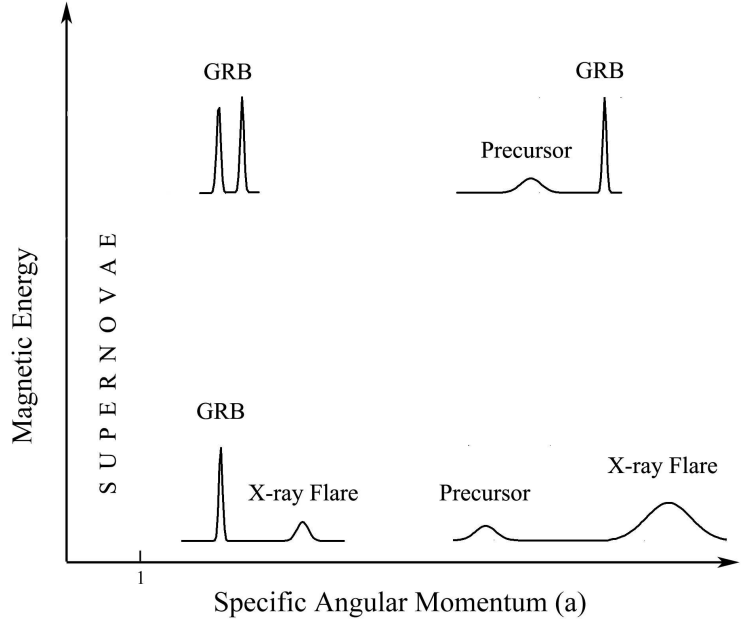}
\caption{Classification of GRBs and surrounding events in terms of the
spinar paradigm. In the case of weak magnetic field and large angular
momentum (the bottom right curve) the first burst is weak (because of the
high centrifugal barrier) and soft. It is followed by slow collapse
(magnetic field is weak), which results in a weak x-ray burst. In the case
of small angular momentum (the bottom left curve) the energy released at the
centrifugal barrier is large and the burst appears as a gamma-ray burst,
whereas the second burst, which corresponds to the collapse of the spinar,
is again weak and soft and shows up as a distant x-ray burst. In the case of
even weaker magnetic field the second flare behaves as an extralong Xray
plateau In the case of stronger magnetic field the flare becomes more like a
gamma-ray burst, its energy increases and the flare itself becomes part of a
gamma-ray burst (the top left curve). As we move rightward, the angular
momentum increases and the first flare loses its energy to become a close
precursor of the second flare, which in the case of a strong magnetic field
becomes a powerful gamma-ray burst. The collapse of a core with the
effective Kerr parametr smaller than unity results in a supernova explosion.}
\label{fig2}
\end{figure}

\clearpage

\begin{figure}
\epsscale{.5}
\plotone{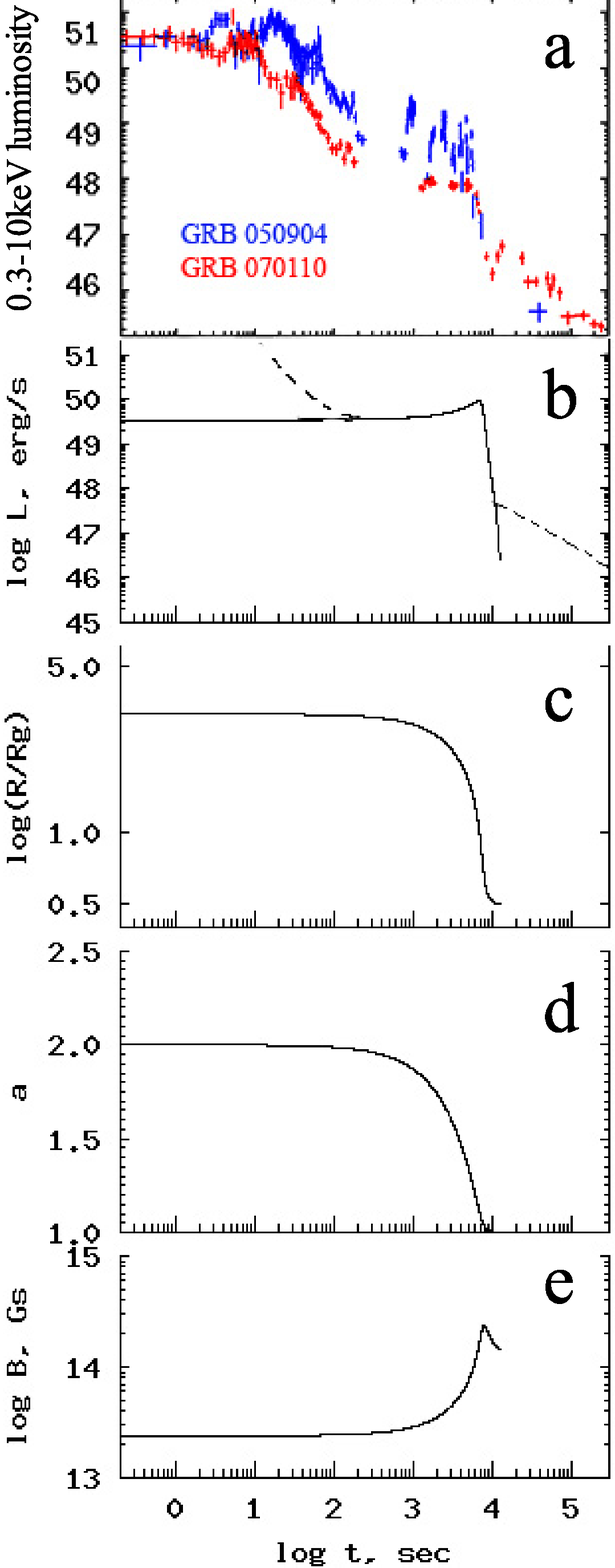}
\caption{Figure 3a Combined Swift (BAT and XRT) light curves of GRB 050904
(z=6.29) and GRB 070110 (z=2.35) in the source rest frame (Troja, E., et al 2007)
Figures 3b-c show the result of the computation of the spinar luminosity,
radius and effective Kerr parameter, respectively. The mean magnetic field
evolution (Figure 3e) is calculated under the approximated general relativistic
model (see eq. 4) }
\label{fig3}
\end{figure}


\begin{thebibliography}{}

\bibitem [][ Bisnovatyi-Kogan, G.S. 1971, Sov. Astron, 14, 652


\bibitem [][ Bogomazov, A.I., V.M. Lipunov, A.V. Tutukov, 2007, ARep 51, 308

\bibitem [][ Chincarini, G., et al., 2007, (astro-ph 0702371).

\bibitem [][ Cusumano et al., 2006, Nature, 440, 164

\bibitem [][ Gehrels, N. et al., 2004, ApJ 611, 1005

\bibitem [][ Gehrels, N. et al.. 2006, Nature 444, 1044

\bibitem [][ Ginzburg, V.L. \& Ozernoy, L.M. 1964, JEPT, 47, 1030

\bibitem [][ Hoyle, F. \& Fowler, W.A. 1963, MNRAS 125, 169

\bibitem [][ Kramer, D. 1984, Class.Quantum Grav., 1, L45

\bibitem [][ Kumar, P. \& Panaitescu, A., 2000, ApJ, 541, L51

\bibitem [][ Lazzati, D. 2005, MNRAS, 357, 722

\bibitem [][ LeBlanc, J.H. \& Wilson, J.R. 1970, ApJ 161, 541

\bibitem [][ Lipunov, V.M., Postnov, K.A. {\&} Prokhorov, M.E., A\&A, 1987, 186, L1

\bibitem [][ Lipunov, V.M. 1992, Astrophysics of Neutron Stars, (Springer-Verlag Berlin New York)

\bibitem [][ Lipunov, V.M. et al., 2006, GCN 5901

\bibitem [][ Lipunova, G.V. 1997, Astronomy Letters, 23, 84

\bibitem [][ Lipunova, G.V. \& Lipunov, V.M. 1998, A\&A, 329, L29

\bibitem [][ Manko, V.S. \& Sibgatullin, N.R. 1992, Class.Quantum Grav., 9, L87



\bibitem [][ Mukhopadhyay, B. 2002, ApJ, 581, 427

\bibitem [][ Ozernoy, L.M. 1966, Soviet Astronomy, 10, 241

\bibitem [][ Ozernoy, L.M. \& Usov, V.V. 1973, Ap\&SS, 25, 149


\bibitem [][ Thorne, K.S., Price, R.H. \& Macdonald, D.A. 1986 Black Holes: The
Membrane Paradigm, (Yale University Press)

\bibitem [][ Troja, E., et al., 2007, (astro-ph 0702220).

\bibitem [][ Tutukov, A., Cherepashchuk, A. 2003, ARep 47, 386

\bibitem [][ Vietri \& Stella, 1998, ApJ, 507, L4

\bibitem [][ Wang, Xiang-Yu \& Meszaros, P. 2007,
(astro-ph 0702441)


\end{thebibliography}
\end{document}